\pdfoutput=1
\documentclass{article}

\usepackage{preamble}

\newcolumntype{d}[1]{D{.}{.}{#1}}
\usepackage[nonatbib,final]{neurips_2023_ml4ps}

\title{19 Parameters Is All You Need:\\Tiny Neural Networks for Particle Physics}

\author{%
  Alexander Bogatskiy\\
  Center for Computational Mathematics\\
  Flatiron Institute, New York, NY, U.S.A.\\
  \texttt{abogatskiy@flatironinstitute.org}\\
  \And
  Timothy Hoffman\\
  Department of Physics, University of Chicago\\
  Chicago, IL, U.S.A.\\
  \texttt{hoffmant@uchicago.edu}\\
  \And
  Jan T.~Offermann\\
  Department of Physics, University of Chicago\\
  Enrico Fermi Institute\\
  Chicago, IL, U.S.A.\\
  \texttt{jano@uchicago.edu}\\
}

\begin{document}

\maketitle

\begin{abstract}
  As particle accelerators increase their collision rates, and deep learning solutions prove their viability, there is a growing need for lightweight and fast neural network architectures for low-latency tasks such as triggering. We examine the potential of one recent Lorentz- and permutation-symmetric architecture, PELICAN, and present its instances with as few as 19 trainable parameters that outperform generic architectures with tens of thousands of parameters when compared on the binary classification task of top quark jet tagging.
\end{abstract}

\section{Introduction}

Particle collisions at the Large Hadron Collider at CERN happen every 25 nanoseconds, producing immense amounts of data that have to be processed in real time. Much of the event filtering is done by the Level-1 trigger~\cite{Buttinger:1456546,ATLAS:2021tnq}, which uses algorithms implemented on FPGAs that need to operate at below-microsecond latency to avoid loss of valuable data. Low-latency tasks include charged particle track reconstruction and energy measurements. Implementing neural networks under such constraints is a significant challenge, however the most recent attempts to do so have finally surpassed their traditional non-ML counterparts. The current state-of-the-art implementations in this area are based on the JEDI-net Graph Neural Network (GNN) architecture, see~\cite{JEDI-net,JEDI22,JEDI23}. The network input data consist of lists of jet constituents, with a certain number of geometric features describing each constituent. 

GNN architectures are inherently permutation-equivariant, providing a significant boost to efficiency and model size by virtue of weight sharing, but no other physical symmetries are necessarily respected. Physics-informed architectures that are inherently equivariant with respect to rotational and Lorentz-boost symmetries have recently shown themselves to provide state-of-the-art performance at tasks such as jet tagging (see e.g.~\cite{Bogatskiy:2020tje, LorentzNet22, KasiePlehn19, DoesItHelp, PELICAN22, PELICAN23}), and they do so despite the relatively small model size. 

In this work we study the current state-of-the-art architecture for top-quark jet tagging, PELICAN \cite{PELICAN23}. It is fully Lorentz-invariant and its permutation-equivariant layers are based on the general higher-order permutation-equivariant mappings introduced in \cite{Maron18,KondorPan}. The full reduction of all relevant symmetries allows small instances of PELICAN with just a few thousand parameters to perform on par with much larger models with hundreds of thousands or even millions of parameters. Moreover, the simplicity of the architecture presents a unique opportunity for explainability and even interpretability. 

Our goal here is to explore the small model size limit of PELICAN and compare it against the previous state-of-the-art (and also Lorentz-equivariant) architecture, LorentzNet \cite{LorentzNet22}. The benchmark task for this comparison is that of top-quark tagging due to the publicly available dataset \cite{KasPleThRu19} and the extensive prior exploration of architectures trained on it \cite{KasiePlehn19}. The input consists of a list of $N$ 4-momenta of jet constituents, $\{p_i\}_{i=1}^N$ which PELICAN reduces to the $N\times N$ array of pairwise Lorentz-invariant dot products, $d_{ij}=p_i\cdot p_j$. Thus the reduced input is an array with one channel. We find that a stripped down version of PELICAN consisting of nothing but two linear permutation-equivariant blocks with just two channels in the hidden layer and exactly one nonlinear activation function in between outperforms generic architectures such as the fully-connected TopoDNN which has 59k parameters \cite{KasiePlehn19}. This model nominally has 26 parameters, but through absorption of multiplicative factors and a simplification of the output layer that number can be effectively reduced to 19. Despite the costly $N^2$ scaling of the memory that PELICAN requires, its symmetric architecture can provide ultra-lightweight networks that can be viably used in low-latency and high-throughput applications.

\section{The original PELICAN architecture}

The original PELICAN architecture consists of an input block which encodes the $N\times N$ array of pairwise dot products $\{d_{ij}\}$, followed by a sequence of so-called $\text{Eq}_{2\to 2}$ permutation-equivariant blocks (the index $2$ indicates the rank of the input and output arrays) that produce transformed $N\times N$ arrays. Each of these blocks consists of a fully-connected ``messaging'' layer that mixes the channels but is shared among all components of the $N\times N$ array, and an ``aggregation'' layer that applies a general linear permutation-equivariant operation that exchanges information between the various components of the array. Finally, a similar $\text{Eq}_{2\to 0}$ block reduces the array to a permutation-invariant (rank 0) scalar, after which an output MLP layer produces the two binary classification weights $\{w_0,w_1\}$. The diagram below summarizes this architecture, see \cite{PELICAN23} for details.
\[
\begin{tikzcd}[column sep=small, row sep=small]
\{d_{ij}\} \arrow[r]& \text{Emb} \arrow[r] & \left[\mathrm{Eq_{2\to 2}}\right]^L \arrow[r] & \mathrm{Eq_{2\to 0}} \arrow[r] & \mathrm{MLP} \arrow[r] & \{w_c\}
\end{tikzcd}
\label{Classifierlayer}
\]
Notably, the aggregation step inside $\text{Eq}_{2\to 2}$, called $\text{LinEq}_{2\to 2}$, applies 15 different operations that provide a basis for the space of all linear permutation-equivariant transformations of rank 2 arrays, which temporarily increases the size of the activation by a factor of 15, marking the peak of PELICAN's memory utilization. This is followed by a trainable linear layer that applies $(C_{\text{in}}\times 15)\times C_{\text{out}}$ weights and adds two biases per channel (one bias added to the entire $N\times N$ array and one only to the diagonal), where $C_{\text{in}}$ and $C_{\text{out}}$ are the number of input and output channels. Similarly, $\text{Eq}_{2\to 0}$ involves only 2 aggregators (total sum and trace), a linear layer of shape $(C_{\text{in}}\times 2)\times C_{\text{out}}$, and one bias per channel.

\section{nanoPELICAN architecture}

In this section we simplify the PELICAN architecture to a single hidden layer and reduce parameters further based on symmetry arguments, which we refer to as nanoPELICAN (nPELICAN). The only two linear symmetric observables that can be constructed from the input dot products (assuming sum-based aggregation that does not explicitly depend on the multiplicity $N$) are $N$, the jet mass $m_J^2=\sum_{i,j} d_{ij}$, and the total mass $\sum_i d_{ii}$. The top-tagging dataset has only massless constituents, so the latter observable is irrelevant. A non-parametric top-tagger based on a simple jet mass cut achieves an AUC of only $90.6\%$. A linear PELICAN, which outputs $p(N)m_J^2+q(N)$ with some learned polynomials $p$ and $q$, cannot far exceed this. 

To this end, we set out to find the smallest and most interpretable modification of PELICAN that is still nonlinear and performs competitively on the top-tagging task. We thus omit the input embedding layer, all messaging layers, and the output MLP, and are left with just two linear equivariant blocks, $\text{LinEq}_{2\to 2}$ and $\text{LinEq}_{2\to 0}$, separated by a single activation function, which we choose to be ReLU. The architecture is summarized in the following diagram:
\[
\begin{tikzcd}[column sep=small, row sep=small]
\{d_{ij}\} \arrow[r] & \text{LinEq}_{2\to 2}^\text{nano} \arrow[r] & \text{ReLU} \arrow[r] & \text{LinEq}_{2\to 0} \arrow[r] & \{w_c\}.
\end{tikzcd}
\label{Classifierlayer}
\]
Here, we also notice that since the array of dot products, $\{d_{ij}\}$, is symmetric, and since the constituents in the top-tagging dataset are massless ($d_{ii}=0$), many of the 15 basis aggregators in $\text{LinEq}_{2\to 2}$ are redundant. We remove 5 aggregators that depend only on the diagonal of the input, and one from each of 4 pairs of aggregators that attain the same value on symmetric inputs. We are left with just 6 aggregators which constitute $\text{LinEq}_{2\to 2}^\text{nano}$. To help with training, each equivariant layer is still preceded by a \texttt{Dropout} layer. Moreover, keeping \texttt{BatchNorm} layers just before the dropout can also help the model converge, meanwhile the extra parameters from these layers can be almost completely absorbed into the linear layers for inference. Namely, the multiplicative weights of \texttt{BatchNorm} can be absorbed into the following $\text{LinEq}_{2\to 2}$, whereas the biases can be either left or absorbed into the biases of $\text{LinEq}_{2\to 2}$ at the cost of turning them into quadratic polynomials of $N$, adding 2 parameters per output channel. Since there are two distinct bias parameters per channel, such a \texttt{BatchNorm} effectively adds $\min\{C_{\mathrm{in}},4C_{\mathrm{out}}\}$ parameters. In the case of $\text{LinEq}_{2\to 0}$ there is only one bias parameter per channel, thus the number of added parameters is $\min\{C_{\mathrm{in}},2C_{\mathrm{out}}\}$.

The only remaining hyperparameter is $C_\text{hidden}$, the number of channels in the hidden layer (between $\text{LinEq}_{2\to 2}$ and $\text{LinEq}_{2\to 0}$). The total number of parameters is then $1\times 6\times C_\text{hidden}+2\cdot C_\text{hidden}+C_\text{hidden}\times 2\times 2+2=12C_\text{hidden}+2$ (ignoring \texttt{BatchNorm}). In addition, since for binary classification only the difference in weights $w_1-w_0$ matters, it is possible to have only 1 output channel, in which case we have $10C_\text{hidden}+1$ parameters. The models presented below produce only one output weight called $w$. Leaving in the two \texttt{BatchNorm} layers effectively adds only 3 new parameters if $C_\text{hidden}>1$ and 2 otherwise. Finally, since we're using the ReLU activation, which is a homogenous function, one more multiplicative factor can be absorbed in each channel. The final number of parameters then is $9C_\text{hidden}+4$ for $C_\text{hidden}>1$ and 12 otherwise.

\section{Top tagging performance}

The top tagging dataset \cite{KasPleThRu19} consists of anti-$k_T$ jets~\cite{Cacciari:2008gp} corresponding with top quarks (signal) and light quarks or gluons (background). It includes up to 200 jet constituents per entry, each represented by a 4-momentum in Cartesian coordinates. A converted version of the dataset that can be directly used with PELICAN can be found at \cite{Gong}. For our models we only use the 80 constituents with the highest transverse momentum $p_T=\sqrt{p_x^2+p_y^2}$, which is typically enough to saturate our network's performance. We follow a training regime almost identical to that in \cite{PELICAN23}, using an Nvidia H100 GPU. The only changes are that we disable weight decay, extend the training to 140 epochs (4 epochs of linear warm-up, 124 epochs of \texttt{CosineAnnealingLR} with $T_0=4$ and $T_\text{mult}=2$, and 12 epochs of exponential decay with $\gamma=0.5$), and increase the batch size to 512. Training took about $30\,\text{ms}$ per batch, and the evaluation took about $23\,\text{ms}$ per batch (including overhead).

\begin{wraptable}{l}{0.6\linewidth} 
    \vspace{-0.6\intextsep}
    \caption{Comparison of tiny top-taggers. Averaged over the top 5 (lowest loss) out of 25 random seeds. Uncertainty given by the standard deviation. $1/\epsilon_B$ is the background rejection at 30\% signal efficiency.}
    \label{tab1}
    \centering
    \begin{small}
    \begin{tabular}{l@{\hspace{2mm}}l@{\hspace{2mm}}l@{\hspace{2mm}}l@{\hspace{2mm}}r}
    \toprule
    Architecture    &   Accuracy    &   AUC         &   $1/\epsilon_B$  &   \# Params \\
    \midrule
    $\text{LorentzNet}_{n_\text{hidden}=3}$  &   0.907(2)    &   0.966(3)    & 174$\pm$44   &   120    \\ 
    $\text{nPELICAN}_{C_\text{hidden}=10}$   &   0.921(1)    &   0.9748(1)   & 327$\pm$20   &   101    \\ 
    $\text{nPELICAN}_{C_\text{hidden}=3}$    &   0.919(1)    &   0.9730(4)   & 256$\pm$12   &   31     \\ 
    $\text{nPELICAN}_{C_\text{hidden}=2}$    &   0.918(1)    &   0.9718(6)   & 243$\pm$18   &   21     \\ 
    $\text{nPELICAN}_{C_\text{hidden}=1}$    &   0.895(1)    &   0.950(2)    &  81$\pm$12   &   11     \\ 
    \bottomrule
    \end{tabular}
    \end{small}
\end{wraptable}
We train several models with $C_\text{hidden}$ ranging from 1 to 10. For comparison, we also train instances of LorentzNet with only one message passing block and the number of channels in the hidden layers set to 3 and the batch size of 512 (no other changes to hyperparameters and training were made). The results are reported in Table~\ref{tab1}. We report the accuracy, the area under the ROC, and the background rejection (inverse false positive rate) at 30\% signal efficiency (true positive rate). For each architecture, we pick the model with the lowest cross-entropy loss out of 10 trained instances initialized with different random seeds.

\begin{figure}[t]
    \vspace{-1em}
    \begin{center}
    \includegraphics[width=0.7\textwidth]{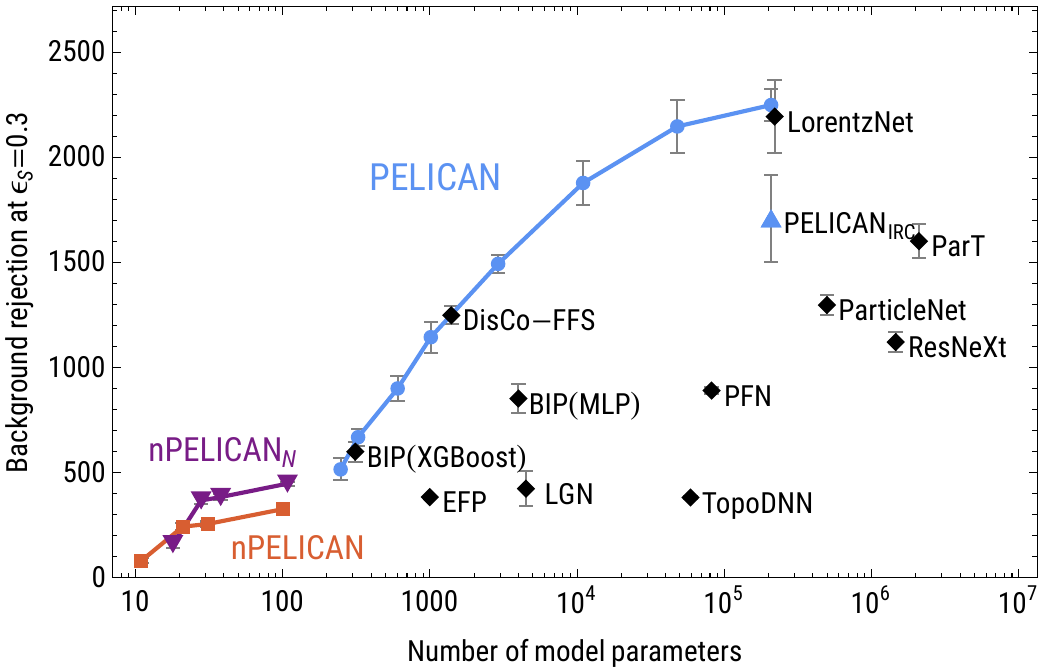}
    \caption{Comparison of top-tagger background rejection performance at signal efficiency $\epsilon_{S} = 0.3$ as a function of the number of parameters in each model considered. Results other than nPELICAN are taken from refs.~\cite{KasiePlehn19, Pearkes:2017hku, EFN, EFP, Bogatskiy:2020tje, Ortner22, Shih22, ParticleNet, ParT, LorentzNet22, PELICAN23}. Note that the curve for the original PELICAN was obtained by varying only the network width, so only the rightmost point is fully optimized.}
    \label{toptag-params}
    \end{center}
\end{figure}

We observe that nPELICAN achieves competitive AUC and background rejection with as few as 2 channels in the hidden layer. In fact, its AUC surpasses that of the fully-connected TopoDNN with 59k parameters \cite{KasiePlehn19} (its average accuracy was $0.929(1)$, AUC $0.964(14)$, and background rejection $424\pm 82$). Moreover, the AUC of nPELICAN with 10 channels (101 parameters) is only about 1\% behind that of ParticleNet (498k parameters) \cite{ParticleNet} and even ParT (2.1M parameters) \cite{ParT}. Meanwhile, LorentzNet with one message-passing block lags far behind nPELICAN with a similar number of parameters. A visual comparison with many existing models is presented in Figure~\ref{toptag-params}.

\begin{wraptable}[9]{l}{0.6\linewidth} 
    \vspace{-\intextsep}
    \caption{Performance of $\text{nPELICAN}_{N}$ -- nPELICAN with $N^\alpha$-scaled aggregators. Metrics defined as in Table \ref{tab1}. \label{tab2}}
    \centering
    \begin{small}
    \begin{tabular}{l@{\hspace{2mm}}l@{\hspace{2mm}}l@{\hspace{2mm}}l@{\hspace{2mm}}r}
    \toprule
    $\text{nPELICAN}_{N}$ width    &   Accuracy    &   AUC         &   $1/\epsilon_B$  &   \# Params \\
    \midrule
    $C_\text{hidden}=10$   &   0.923(1)    &   0.9764(1)   & 448$\pm$10   &   108   \\ 
    $C_\text{hidden}=3$    &   0.9214(3)    &   0.9752(2)  & 384$\pm$16   &   38    \\ 
    $C_\text{hidden}=2$    &   0.9200(3)   &   0.9745(1)   & 368$\pm$17   &   28    \\  
    $C_\text{hidden}=1$    &   0.902(2)    &   0.960(2)    & 150$\pm$16   &   18    \\ 
    \bottomrule
    \end{tabular}
    \end{small}
\end{wraptable}

Interestingly, at such low depth the removal of fully connected messaging layers from PELICAN actually improved the performance. The element of the original network that can boost nPELICAN's performance the most with only a few new parameters is the $N$-dependent scaling of aggregators. In our tests, replacing sum aggregation with means led to very low performance of nPELICAN, however enabling PELICAN's original flexible scaling of the means by an extra factor of $N^{\alpha}/{\bar N}^\alpha$ turns out to be very beneficial, see Table \ref{tab2}.

\section{Interpreting nanoPELICAN}

Considering the extremely low complexity and relatively high performance of nPELICAN, there is high potential for a full interpretation of the model. Before attempting to interpret the weights, it is crucial to minimize any redundancies. In particular, since \texttt{ReLU} is a homogenous function, one multiplicative factor from the weights in $\text{LinEq}_{2\to 0}$ can be absorbed into the weights and biases of $\text{LinEq}_{2\to 2}$ in each channel of the hidden layer. For the model with $C_\text{hidden}=2$ this means that the number of parameters can effectively be reduced to 19. Explicitly, the model can be written analytically as
\begin{multline}
    w= b^{2\to 0}+\sum_{h=1}^{C_\text{hidden}}c^{2\to 0}_{0h} \frac{1}{\bar{N}^2}\sum_{i,j} \text{ReLU}\left(\sum_{b=1}^{6}c^{2\to 2}_{bh}\text{Agg}_b(d)_{ij}+b^{2\to 2}_h+b^{2\to 2}_{\text{diag},h}\delta_{ij}\right)+\\
    +\sum_{h=1}^{C_\text{hidden}}c^{2\to 0}_{1h} \frac{1}{\bar{N}}\sum_{i=j} \text{ReLU}\left(\sum_{b=1}^{6}c^{2\to 2}_{bh}\text{Agg}_b(d)_{ij}+b^{2\to 2}_h+b^{2\to 2}_{\text{diag},h}\delta_{ij}\right).
\end{multline}
Here, $w$ is the output score (the jet is tagged as a top quark if $w>0$); $c^{2\to 2}$, $b^{2\to 2}$, and $b^{2\to 2}_\text{diag}$ are the weights and biases of $\text{LinEq}_{2\to 2}$; $c^{2\to 0}$ and $b^{2\to 0}$ are the weights and the bias of $\text{LinEq}_{2\to 0}$; index $b$ enumerates the $6$ aggregators of $\text{LinEq}_{2\to 2}$; index $h$ enumerates the channels in the hidden layer. $\bar{N}$ is a hyperparameter that is used to control the magnitude of the sums over constituents, here set to be 49 (it is similarly used inside the aggregators $\text{Agg}_b$).

Therefore the ReLU effectively sets a linear constraint on the dot products, and the output takes a sum over only those pairs $(i,j)$ that satisfy the constraint. More explicitly, denoting the jet momentum by $J=\sum_i p_i$, the argument of the ReLU is a linear combination of relative masses ($m_{ij}^2=-(p_i-p_j)^2=2d_{ij}$), jet-frame masses $p_i \cdot J$, $p_j \cdot J$, the jet mass $m_J^2=\sum_{ij}d_{ij}$, and a constant term. In addition, we found that these parameters are stable across multiple random initializations, indicating that they can be directly interpreted as unique physical constraints that encode Lorentz-invariant quantities such as the top quark mass, which we intend to elucidate in future work.

\section{Conclusions}
We have presented nPELICAN, a miniaturized a version of the PELICAN architecture, which is both surprisingly performant relative to much larger networks and can be rewritten simply as a constraint on Lorentz invariant quantities with a single ReLU activation function. This represents a novel development in interpretability for neural networks in particle physics, and gives hope for the interpretability of much larger networks. Future studies will exploit the stability of the nPELICAN parameters to determine their dependencies on features of the training data such as jet energies and particle masses,
as well as relating these parameters to traditional, discriminating kinematic variables for jet-tagging, such as jet constituent multiplicity, subjet multiplicity~\cite{Thaler:2010tr} and jet shapes~\cite{Ellis:1992qq}.
The code can be found at \url{https://github.com/abogatskiy/PELICAN-nano}.

\setlength\bibhang{0pt} 
\setlength\bibitemsep{2pt} 
\setstretch{0.9} 
\setquotestyle{english}
\printbibliography[heading=bibintoc]
\setstretch{1} 

\end{document}